\documentclass[conference]{IEEEtran}
\IEEEoverridecommandlockouts
\usepackage[numbers,sort,compress]{natbib}
\usepackage{xfrac}
\usepackage{booktabs}
\usepackage{url}

\usepackage{subfigure}
\usepackage{amsmath,amssymb,amsfonts}
\usepackage{algorithmic}
\usepackage{graphicx}
\usepackage{textcomp}
\usepackage{xcolor}
\usepackage{nicefrac}

\def\BibTeX{{\rm B\kern-.05em{\sc i\kern-.025em b}\kern-.08em
    T\kern-.1667em\lower.7ex\hbox{E}\kern-.125emX}}

\usepackage{makecell}

\usepackage{pgfplots}
\pgfplotsset{compat=newest}

\begin{document}

\newtheorem{theorem}{Theorem}
\newtheorem{lemma}{Lemma}
\newtheorem{definition}{Definition}

\renewcommand{\vec}[1]{\mathbf{#1}}
\newcommand{\ve}[1]{\boldsymbol{#1}} % for greek letters

\newcommand{\diag}{\mbox{diag}}
\newcommand{\tr}{\mbox{tr}}
\newcommand{\F}{{\cal{F}}}
\newcommand{\Om}{\Omega}
\newcommand{\om}{\omega}
\newcommand{\Pp}{ \mathbb{P} }
\newcommand{\R}{\mathbb{R}}
\newcommand{\Z}{\mathbb{Z}}
\newcommand{\E}{\mathbb{E}}
\newcommand{\En}{\mathbb{E}_n}
\newcommand{\Et}{\tilde{\mathbb{E}}}
\newcommand{\Etn}{\tilde{\mathbb{E}}_n}
\newcommand{\borel}{ {\cal{B}}(\mathbb{R}) }
\newcommand{\pspace}{(\Om, \F, \Pp)}
\newcommand{\Pt}{\tilde{\mathbb{P}}}

\newcommand{\Var}{\mbox{Var}}
\newcommand{\Cov}{\mbox{Cov}}
\newcommand{\RR}{\varrho _{1,2}}

%% Citation package
%\usepackage{natbib} % to use \citet
%\bibliographystyle{abbrvnat}
%\setcitestyle{authoryear,open={(},close={)}}

% \title{Quadratic optimal portfolio choice under distribution misspecification}
\title{The efficient frontiers of mean-variance portfolio rules under distribution misspecification}
% Portfolio Performance of Shrinkage Estimators Under Distribution Misspecification}
% miss-specification, degeneracy, ambiguity distributional ambiguity? -- looking for a catchy phrasing the original title was pretty good

\author{\IEEEauthorblockN{Andrew Paskaramoorthy}
\IEEEauthorblockA{\textit{School of Computer Science and Applied Mathematics} \\
\textit{University of the Witwatersrand},\\
Johannesburg, South Africa \\
andrew.paskaramoorthy@wits.ac.za}
\and
\IEEEauthorblockN{Tim Gebbie}
\IEEEauthorblockA{\textit{Department of Statistical Sciences} \\
\textit{University of Cape Town},\\
Cape Town, South Africa \\
% email address or ORCID
orcid.org/0000-0002-4061-2621 \\
tim.gebbie@uct.ac.za
\and
\IEEEauthorblockN{Terence van Zyl}
\IEEEauthorblockA{\textit{Institute for Intelligent Systems} \\
\textit{University of Johannesburg},\\
Johannesburg, South Africa \\
tvanzyl@uj.ac.za
}
}}

\maketitle

\begin{abstract}
Mean-variance portfolio decisions that combine prediction and optimisation have been shown to have poor empirical performance. Here, we consider the performance of various shrinkage methods by their efficient frontiers under different distributional assumptions to study the impact of reasonable departures from Normality. Namely, we investigate the impact of first-order auto-correlation, second-order auto-correlation, skewness, and excess kurtosis. We show that the shrinkage methods tend to re-scale the sample efficient frontier, which can change based on the nature of local perturbations from Normality. This re-scaling implies that the standard approach of comparing decision rules for a fixed level of risk aversion is problematic, and more so in a dynamic market setting. Our results suggest that comparing efficient frontiers has serious implications which oppose the prevailing thinking in the literature. Namely, that sample estimators out-perform Stein type estimators of the mean, and that improving the prediction of the covariance has greater importance than improving that of the means. 
\end{abstract}


\section{Introduction}

% 1. THE SETTING
\citet{markowitz1952} defined the portfolio selection problem for an investor who seeks to maximise the expected return of his portfolio whilst minimising its variance. The expected returns and covariances of asset returns are assumed to be known to the investor, and asset returns are assumed to be Independently and Identically Distributed (IID). Thus, no statistical inference is required and the optimal decision rule is a function of the parameters for the distribution of returns. Under these assumptions, the mean-variance objective function represents the long-run performance of repeated application of the optimal decision rule. This framework necessarily has strong distributional assumptions. 

% 2. THE KEY LITERATURE SUMMARY
In practice mean-variance portfolios have been shown to have poor empirical performance. Reasons for the poor performance was first associated with the poor predictive performance of sample estimators and the sensitivity of the optimal solution to estimation errors. Consequently, resolving the problems associated parameter uncertainty became a central issue in the application of mean-variance portfolio theory. 

Amongst the main solutions initially proposed were the development of shrinkage estimators to reduce estimation error \cite{BM2009} -- at first with relatively strong limiting assumptions about the distribution of the data. Shrinkage estimators are closely related to Bayesian estimators since the shrinkage target is often used to encode the investor's prior information \cite{black1992global}. Bold attempts were made to shift towards a Bayesian perspective and investigate portfolio rules for different prior distributions. \citet{klein1976effect} investigate non-informative priors. \citet{frost1986empirical} and \citet{jorion1986bayes} present Empirical-Bayes estimators for parameters that outperform sample estimators, measured by expected utility. 

% However, in the end the performance improvement of these modifications was disappointing and at times insignificant \cite{BM2009}.  

Some of the thinking then shifted towards a more generalised approach that could incorporate estimation errors in the specification of the portfolio selection problem itself; perhaps by using a Decision-Theoretic (DT) approach. The resulting decision rules consist of inference and decision stages, and resemble the ``plug-in'' approach used but in conjunction with shrinkage estimation. More recently, \citet{kan2007optimal} and \citet{tu2011markowitz} use a Decision-Theoretic framework to design portfolio rules that combine simpler portfolio rules. Studies using the DT framework seemed to demonstrate, either analytically or via simulation, the out-performance of the respective portfolio decision rule over the sample estimator plug-in rule in terms of expected utility.

However, in a landmark empirical survey, \citet{BM2009} reported that various seminal decision rules often perform poorly on historical market data-sets. The poor empirical performance of these methods may be due to sampling variation, but it may arise as a result of the discrepancy between the assumed and actual return generating process. Alternatively, the poor reported performance may be a consequence of the evaluation methodology. 

% 3. THE IDEA

In practice, one key problem is that the assumption of an IID return generating process is simply untrue. Being IID is assumed to ensure mathematical tractability of the problem analysis. It is well known that empirical properties of data deviate from this assumption \cite{cont2001empirical}. However, one can still identify at least four subordinate reasons, and then try and exploit these to ameliorate the failure of the empirical performance to converge to the theoretical performance: parameter mis-specification, distribution mis-specification, inadequate sampling, and the role of portfolio constraints (See Table \ref{tab:reasons}). 

\begin{table}[h]
    %\scriptsize
    %\centering
    \caption{Key causes for the lack of consistency between empirical and theoretical portfolio selection decisions.}
    \label{tab:reasons}
    \begin{tabular}{p{8.5cm}}
    \toprule
        {\bf Parameter Mis-specification}: In simulation studies the performance is measured using particular combinations of parameters for some assumed distribution. This is typically chosen to be a Multivariate Normal distribution; it is possible that the simulated performance may be poor under different parameter or distribution choices.\\
    \midrule
        {\bf Distribution Mis-specification}: Analytical studies demonstrate dominance across the parameter space, but again for a single assumed global distribution. Empirical returns are known to deviate from the Normal distribution, and this may adversely affect the performance of the estimators.\\
    \midrule
        {\bf Inadequate Samples}: the theoretical performance is measured by expected utility which represents the average realised performance from repeatedly applying the decision on independent samples. Given the length of financial time series, convergence may not be practically possible. \\
    \midrule
        {\bf Imposition of Constraints}: some of the studied decision rules are developed for the case where a risk-free asset is included in the investible set. However, \cite{BM2009} studies the case where there is no risk-free asset. This requires that the original portfolio rules are transformed by a seemingly benign scaling to ensure weights sum to 1. However, this scaling modifies how errors are amplified in the optimisation\cite{PASKARAMOORTHY2021102065}. \\ 
    \bottomrule
    \end{tabular}
    % fix the font
    
\end{table}

The main assumptions employed in the development and analysis of the portfolio decision rules is still that returns are generated IID, and then often sampled from a multivariate Gaussian distribution. We analyse the performance of these methods under different distributions to study reasonable departures from Normality. Namely, we investigate the impact of: i.) first-order autocorrelation, ii.) second-order autocorrelation, iii.) skewness, and iv.) excess kurtosis. 

Furthermore, an important but previously unrecognised key issue in performance measurement is that decision methods are traditionally compared for a particular specification of the risk-aversion parameter \cite{BM2009}. In theory, the risk-aversion parameter reflects the trade-off between the expected return and variance and is specified prior to portfolio construction. However, in application, the risk-aversion parameter can be treated as a hyper-parameter that is optimised as part of the portfolio construction process in order to achieve the desired out-of-sample performance \cite{boyd2017multi}. Thus, rather than comparing portfolios from each decision rule for particular levels of the risk-aversion parameter, as done in the standard Decision Theoretic framework, we compare the entire efficient frontiers of the decision rules. 

% 4. WHAT WE WILL DO
Our main contributions are to show that comparing portfolio decision rules on the basis their efficient frontiers, reveals insights which run counter to the prevailing understanding in the portfolio selection literature. In particular, we show that:
\begin{enumerate}
    \item the plug-in rule has superior \textit{frontier performance} over the Bayes-Stein mean estimator.
    \item Improving the estimation of the covariance matrix results in greater performance gains than improving the estimation of the mean.
\end{enumerate}
Our investigations reveal that these findings are robust to departures from the assumption of an IID multivariate Gaussian data-generating process (DGP). Furthermore, our results show that changes in the DGP can have substantial impact on the performance of all mean-variance based decision rules, but that the relative performance of the decision rules largely remains the same. 

The rest of this paper is organised as follows. In Section \ref{sec:PPDR} we discuss the evaluation of portfolio decision rules. In Section~\ref{sec:psupu}, we outline the mean variance portfolio selection framework and briefly describe the portfolio selection rules investigated in our study. In Section~\ref{sec:method} we describe our methodology, in Section~\ref{sec:results} we present and discuss our results, and in Section \ref{sec:conclusion} we conclude.  

\section{Performance Measurement of Portfolio Decision Rules}\label{sec:PPDR}

Different decision rules can be constructed using different estimators for the mean and variance parameters by modifying the optimisation problem to account for parameter uncertainty. Since the data is random, the utility of a decision rule is a random variable, but with a distribution that depends on its particular specification and the true data-generating process. Different portfolio decision rules are typically compared by evaluating some objective function that is taken to represent an expected utility $\E_\mathcal{S}[U(\hat{\ve \omega})]$ with portfolio weight vector $\ve \omega$, and the expectation is taken over the random samples, $\mathcal{S}$. In theory, if a decision rule is applied repeatedly over independent samples, its realised performance will converge to $\E_\mathcal{S}[U(\hat{\ve \omega})]$. In practice, empirical performance may differ substantially from $\E_\mathcal{S}[U(\hat{\ve \omega})]$ owing to differences in the data-generating process and whether there are an adequate number of representative and independent samples. 

Except for some simple cases, the expected utility for a particular decision rule is not readily analytically tractable, and hence are computed by simulation. In the finance literature, the standard procedure for performance analysis is to generate IID data from a multivariate Gaussian distribution, where parameters are specified using sample estimates from a historical dataset. Then, for each sample, portfolios are constructed using the decision rule for a particular risk-aversion level, which are then used to evaluate the expected utility using the true parameters. Typically, evaluation will occur at multiple levels of risk-aversion, resulting in performance measurements for different portfolios along the efficient frontier. Furthermore, results are typically tabulated, presenting expected utility for different simulation parameters and for different risk aversion levels. For example, in Table \ref{tab:expU} we tabulate the percentage loss in expected utility for the different decision rules considered. 

One convenient way of thinking about this is to use an exponential objective $U(\ve \omega) \propto e^{-\gamma \ve \omega' \vec r}$ for some realisation $\vec r$ from the DGP, and then take a Taylor expansion:
%because for some $\epsilon \sim N(\mu,\sigma^2)$ we then have that $\E[e^{-\gamma \epsilon}] = e^{-\gamma \mu + \frac{\gamma^2}{2} \sigma^2}$:
\begin{align}
    \max_{\ve \omega} \E_t\left[{- \frac{1}{\gamma}e^{- \gamma \ve \omega' \ve r_t}}\right] \approx \max_{\ve \omega} \left\{ { \ve \omega' \E_t[\vec r] -  \frac{\gamma}{2} \ve \omega' \Sigma \ve \omega} \right\} + {\cal O}(3) \nonumber 
\end{align}
so that with the assumption of Normality this reduces exactly to the mean-variance problem. For deviations from Normality the higher moments should be included in the objective function if one wants to retain the exponential form of the utility. 

However, this is not what is typically done, it is usual to retain the problems mean-variance formulation but use data from the real DGP arguing that the differences between the empirical and theoretical performance are due to parameters mis-estimation. If the performance of a portfolio is measured by the value of the mean-variance objective function, but under the true distribution of returns, it should be surprising if the theoretical and empirical results could align along the entire efficient frontier in an equivalent manner. 

For this reason we compare the efficient frontier as whole, as this better reflects how portfolio selection occurs in practice. To do this, we simply compare the mean return of each decision rule for a given variance. Thus, we allow the risk-aversion parameter to vary between decision rules. This is in-line with how portfolios are constructed in practice where the risk-aversion parameter, as well as penalty parameters for other soft constraints in the optimisation, are treated as hyper-parameters which are selected on the basis of backtest performance. It is worth emphasising that, in practice, investors are far more concerned about realised risk and return rather than an ex-ante measure of this trade-off. 

\section{Portfolio Selection Under Parameter Uncertainty} \label{sec:psupu}
A mean-variance investor will construct a portfolio using a
decision rule comprised of a \textit{prediction} and then a \textit{optimisation} step. In the prediction step, the investor will use the available data to determine estimates for the means $\ve \mu$ and covariances $\Sigma$ of returns of the investible assets over the investment period. In the optimisation step, the investor can use the predictions to select a mean-variance portfolio using a quadratic optimisation with a linear constraint:
\begin{align}\label{eq:MVO}
    \operatorname*{argmax}_{\ve \omega} \left\{ { \ve \omega'\ve  \mu - \cfrac{\gamma}{2} \ve \omega'\Sigma \ve \omega }\right\} \quad \mbox{s.t} \quad \ve \omega'\vec 1 = 1.
\end{align}
where $\omega$ is the vector of portfolio weights.
In many theoretical studies, a \textit{full-investment} constraint is commonly imposed which requires that portfolio weights sum to one: $\ve \omega'\vec 1 = 1$. Under this constraint, the set of mean-variance optimal portfolios are given by \cite{lee2000theory}:
\begin{align}\label{eq:EF}
    \ve \omega^{*} = \left(1 - \frac{\vec 1'\Sigma^{-1}\ve  \mu }{\gamma} \right) \frac{\Sigma^{-1}\vec 1}{\vec 1'\Sigma^{-1}\vec 1} + \left(\frac{\vec 1'\Sigma^{-1}\ve  \mu }{\gamma}\right) \frac{\Sigma^{-1}\ve  \mu }{\vec 1'\Sigma^{-1}\ve  \mu }
\end{align}
The optimal solution is a convex combination of two portfolios: the global minimum variance portfolio (GMV) and the portfolio that maximises the risk-adjusted return (MRAR). This result is known as the two fund separation theorem (See \citet{ingersoll1987}). The weighting given to each of these portfolios depends on the estimated quantity $\vec 1'\hat{\Sigma}^{-1}\hat{\ve \mu}$ and the risk aversion parameter $\gamma$. By varying the risk-aversion parameter, we attain the full set of Pareto-efficient portfolios, which is more commonly referred to as the \textit{efficient frontier}. 

\subsection{Portfolio Decision Rules} \label{subsec:pdr}
Since the mean and variance parameters can be modifying for the optimisation problem to account for parameter uncertainty,  different decision rules will construct different GMV and MRAR portfolios, and hence different efficient frontiers, for a given set of data. We consider the efficient frontiers of some of some seminal shrinkage estimators presented in Table \ref{tab:estimators}, which are described in detail below.

\begin{table}[h]
    \centering
    \caption{The selection of decision rules considered in this work.}
    \label{tab:estimators}
    \begin{tabular}{cl}
    \toprule
    Section & Approach \\
    \bottomrule\toprule
    \ref{sssec:bsmeans} & Bayes-Stein (BS) decision rule \cite{jorion1986bayes} \\
    \ref{sssec:scov} & Linear Shrinkage (LS) Estimator for Covariance \cite{ledoit2003improved}\\
    \ref{sssec:nlincov} & Nonlinear Shrinkage (NS) estimator for Covariance \cite{ledoit2017nonlinear,ledoit2020analytical} \\
    \ref{sssec:sbayes} & Bayesian ``Data-and-Model" decision rule \cite{pastor2000portfolio}\\
    \bottomrule
    \end{tabular}
\end{table}

\subsubsection{Bayes-Stein Shrinkage Portfolio} \label{sssec:bsmeans}
The Bayes-Stein shrinkage portfolio adopts an Empirical-Bayesian perspective on portfolio returns, where parameters in the prior distribution of means are estimated from the data \cite{jorion1986bayes}. The Bayes-Shrinkage estimator is derived by maximising the investors utility function using the predictive density of the returns which results in the following estimator for the means:
\begin{equation}
            \hat{\ve \mu}_{_\text{b}} = \left(1-\hat{\phi_b}\right)\hat{\ve \mu}_{_\text{s}} + \hat{\phi_b}\vec 1\hat{ \mu}_{_\text{0}} \\
\end{equation}
where $\ve \mu_{_\text{b}}$ is the Bayes-Stein estimator, $\ve \mu_{_\text{s}}$ is the sample estimator of the means, and there is a weight factor $\hat \phi_b$: 
\begin{equation}
    \hat{\phi}_b = \frac{n+2}{\left(n+2\right) + \left(\hat{\ve \mu}_{_\text{s}} - \vec 1\hat{\mu}_{_\text{0}}\right)^{\top} T \Sigma^{-1}\left(\hat{\ve \mu}_{_\text{s}} - \vec 1 \hat{ \mu}_{_\text{0}}\right)  }.
\end{equation}
Here, $\hat{\mu}_{_\text{0}}$ is the observed average return of the minimum variance portfolio, $n$ is the number of assets, and $T$ is the length of the estimation window. Although unknown, the covariance $\Sigma$ matrix of returns is not modelled within the Bayesian framework, and is simply replaced with a sample estimate: $\hat \Sigma_{_\text{s}}$. In the predictive density of returns, the covariance matrix accounts for the estimation risk and is given by:
\begin{align}
    \hat{\Sigma}_{_\text{b}}  = \Sigma\left(1 + \frac{1}{T + \hat{\phi}_b}\right) + \frac{\hat{\phi}_b}{T\left(T+1+\hat{\phi}_b\right)} \frac{\vec 1 \vec 1'}{\vec 1'\Sigma^{-1}\vec 1}
\end{align} \\
To construct a portfolio, the estimators $\hat{\mu}_{\text{bs}}$ and $\hat{\Sigma}_{\text{bs}}$ are plugged into the optimal solution Equation~\ref{eq:EF}.

\subsubsection{Linear Shrinkage Covariance Estimator} \label{sssec:scov}
The Linear Shrinkage covariance estimator combines a shrinkage target, often chosen to be a factor model, a constant correlation covariance matrix, or the identity matrix, with the sample covariance matrix. The optimal shrinkage intensity is derived through minimising a loss function that is defined by the Frobenius norm of the difference of the estimator and the population covariance matrix. Asympototic arguments are used to derive an estimator for the optimal shrinkage intensity. The shrinkage estimator is given by:
\begin{align}
    \hat{\Sigma}_{_\text{l}} = \hat{\phi_{l}}\mathbf{\hat{F}} + \left(1-\phi_l\right)\hat{\Sigma}_{_\text{s}}
\end{align}
where the shrinkage intensity is calculated by:
\begin{align}
    \hat{\phi}_l = \frac{\hat{\pi}-\hat{\rho}}{T\gamma}
\end{align}
where $\hat \pi$ is sum of the asymptotic variances in the sample covariance matrix, $\hat \rho$ the sum of the asymptotic covariances between asset, and $\gamma$ the squared difference between sample and shrinkage target. In practice, this is modified to ensure that it lies in  $[0,1]$:
\begin{align}
    \hat{\phi}_l = \max \left\{ {0, \min \left\lbrace { \frac{\hat{\pi}-\hat{\rho}}{T\gamma}, 1 } \right \rbrace }\right\}.
\end{align}
Here $\hat \pi$ is an estimator for the asymptotic variance of the sample covariance matrix, $\hat \rho$ is an estimator for the asymptotic covariance between the shrinkage target and sample covariance matrix, risk aversion is denoted by $\gamma$, and sample length by $T$.\\
In our implementation, we construct portfolios using the LS estimator for the covariances with the sample estimator for the means.

\subsubsection{Bayesian ``Data and Model" Portfolio}\label{sssec:sbayes}
Factor models are often used to model asset returns and typically have the form:
\begin{align}
    \vec r_{t} = \ve \alpha_i + B \vec r_{_B,t} + \ve \epsilon_{t}
\end{align}
where $\vec r_{t}$ denotes the vector of risky asset returns at some time $t$, $\vec r_{_B,t}$ are the returns on the $K$ factor portfolios, and $\ve \epsilon_{t}$ is a random error term which follows the usual regression restrictions, and $\ve \alpha$ represents average returns in excess of the risk factor portfolios, and $B$ is the matrix of loadings to each factor. The model implies a factor covariance matrix $\Sigma_{_B}$ of dimension $K$.

% with confidence captured by a vector of standard deviations $\ve \sigma$. 
An investor expresses his prior belief in the asset pricing model with the restriction that $\ve \alpha= \vec 0$, with confidence expressed as a scaling (denoted $\tau$) of the sample covariance matrix $\tau\Sigma$. The quantity $\bar B$ represents the estimates of the factor loadings under the restriction $\ve \alpha= \vec 0$ (full confidence in the asset pricing model), whilst $\hat B$ are the factor loading estimates without this restriction (no confidence in the asset pricing model).

Wang \cite{wang2005shrinkage} shows how this prior information can be combined with sample information to produce the following estimators of the mean given the restricted and unrestricted estimated factor loadings, $\bar B$ and $\hat B$ respectively, for some weighting factor $\phi_d$:
\begin{equation}
    \hat{\ve \mu}_{_\text{d}} = \hat{\phi}_{d} \left( {\bar B \hat{\ve \mu}_{_\text{B}}} \right) + \left( 1 - \hat{\phi}_d \right)\hat{\ve \mu}
\end{equation}
where the covariance matrix estimator is:
\begin{align}
    \hat{\Sigma}_{_\text{d}} = \frac{T+1}{T-K-2} F \hat{\Sigma}_{_B} F' + \frac{T}{T-n-1} D E 
\end{align}
where matrices F, D and E are
\begin{align}
F &= \left({ \hat{\phi_d}\bar{B} + (1-\hat{\phi_d})\hat{B}}\right), \\
D &= \left(\hat{\phi_d}\bar{\delta} + (1-\hat{\phi_d})\hat{\delta} \right), \\
E &= \left(\hat{\phi_d}\bar{\Sigma}_e + (1 - \hat{\phi_d})\hat{\Sigma}_e\right).
\end{align}
The resulting estimator combines the asset pricing model and the sample data, where the shrinkage weighting factor $\hat{\phi}_{d}$ is determined by the relative confidence in the asset pricing model:
\begin{align}
    \hat{\phi}_{d} = \frac{1+\hat S^2}{T \tau + 1 + \hat S^2}
\end{align}
where $\hat{S}^2 = \hat{\ve \mu}' \hat{\Sigma}^{-1}_{_\text{b}} \hat{\ve \mu}$ and represents the  square of maximum Sharpe Ratio attainable on the frontier spanned by the factor portfolios. As before, The sample length is given by $T$.\\

\subsubsection{Nonlinear Shrinkage Covariance Estimator}  \label{sssec:nlincov}
The nonlinear shrinkage covariance estimator shrinks the spectrum of the eigenvalues whilst keeping eigenvectors fixed. Shrinkage is described as nonlinear since each eigenvalue has a different shrinkage intensity. The nonlinear shrinkage estimator is described by:
\begin{equation}
    \hat{\Sigma}_{_\text{n}} = U \hat{\Lambda} U' \quad
\end{equation}
where the adjusted eigenvalues are:
\begin{equation}
\hat{\Lambda} = \mathrm{diag} \left( \hat{\phi_n}(\lambda_1), \dots, \hat{\phi_n}(\lambda_N) \right).
\end{equation}
The shrinkage function for each eigenvalue when $x \neq0$ is:
\begin{equation}
    \hat{\phi}_n(x) = \frac{x}{\left({\pi\hat{c}x\hat{f}(x)}\right)^2 + \left({1 - \hat{c} - \pi\hat{c}x\mathcal{H}_f(x)}\right) }
\end{equation}
when $x=0$ but $\hat{c} > 1$ we then use:
\begin{equation}
           \hat{\phi}_n(x) =   \frac{1}{\pi(\hat{c}-1)\mathcal{H}_{\underline{f}}(0)}. 
\end{equation}
where $\hat{c}=N/T$ denotes an estimate of the limiting concentration ratio, $\mathcal{H}$ is the Hilbert transform, $f$ is the limiting density of the sample eigenvalues when $x>0$, whilst $\underline{f}$ is the limiting density when $c>1$.
As for the linear case, we use a sample mean estimator with the nonlinear shrinkage covariance estimator to construct portfolios in our implementation.

\section{Methodology}\label{sec:method}

Our aim is to compare the true efficient frontier with the expected frontiers of the respective decision rules, under different DGPs. Using a simple Monte-Carlo experiment, we simulate the distribution of the frontiers from which we can calculate the average frontier for a particular decision rule, under a specific DGP. Using the simulation parameters for the DGP, we calculate the true frontier for comparison.

\subsection{Data}
We use the Fama-French (FF) monthly dataset of ten industry portfolios to specify the simulation parameters for each of the data-generating processes considered in our study. This dataset contains monthly value-weighted returns of ten industries over the period July 1927 to October 2019. From this dataset, we use the 120 months of data. 

\subsection{Data Generating Processes}
We consider the data-generating processes given in Table \ref{tab:DGP}. The simulation parameters are specified such that the mean and covariance of simulated returns is the same in each DGP, and are specified using the sample estimates of the Fama-French dataset.

\begin{table}[h]
    \caption{Data Generating Processes (DGP) considered in this work.}
    \label{tab:DGP}
    \centering
    \resizebox{\columnwidth}{!}{%
    \begin{tabular}{clc}
    \toprule
    Distribution & Parameters & Deviation  \\
    \bottomrule\toprule
    Multivariate Gaussian (MVG) & $\ve \mu, \Sigma$ & \- \\
    Multivariate Student-t (MVT) & $\nu, \ve \mu, \Sigma$ & Excess kurtosis \\
    Multivariate Skew-Normal (MVSN) & $\ve \mu, \Sigma, \gamma_1$ & Skewness \\ 
    MVG with AR(1) errors & $\ve \mu, \Sigma, \phi$ & First-order autocorrelation \\
    MVG with GARCH(1,1) errors & $\ve \mu, \Sigma, \ve \alpha, \ve \beta$ & Second-order autocorrelation \\
    \bottomrule
    \end{tabular}}
\end{table}

We use a dirty estimate of the the degrees of freedom parameter $\nu$ for the MVT distribution, which is made using the sample average of the degrees of freedom from the univariate fits of each asset. Note that the $\Sigma$ parameter in the multivariate student's-t distribution is a parameter and not the actual covariance, which is given by $\dfrac{\nu}{(\nu-2)}\Sigma$ where $\nu$ are the degrees of freedom. To determine this parameter, we first specify the covariance using sample estimate from the FF data. Then, substituting in our estimate of $\nu$, we calculate the corresponding $\Sigma$. 

By using the centered parameterisation of the MVSN distribution, we specify its parameters using the sample counterparts. Whilst the MVSN distribution has been used in the analysis of financial returns (see \cite{adcock2015skewed}), we note that it is constrained to a maximum amount of multivariate skewness \cite{azzalini2013skew}, which is less than the sample estimate of our dataset. Thus, we scale the sample skewness vector by a factor of $0.72$ to specify the MVSN skewness parameter $\gamma_1$. 

Auto-correlation is typically present only in high-frequency data, and accordingly we find that the autocorrelation in our dataset is not statistically significant. However, for illustrative purposes we scale the sample auto-correlation estimates such that the cross-sectional average auto-correlation is $\bar{\phi} = -0.15$, which we use as parameters in individual AR(1) return processes. We simulate data $\tilde{\vec x}$ from these AR(1) processes which are then transformed using the square root of covariance matrix of the returns and then added to the mean $\vec \mu$ in order to simulate return $\tilde{\vec r}$:
\begin{align}
    \tilde{r} = \Sigma^{1/2}\tilde{\vec x} + \vec \mu
\end{align}
For simplicity, we do not allow cross-correlation at non-zero lags.

Lastly, we add second-order auto-correlation to the MVG process by simulating data from univariate GARCH(1,1) models that are fitted to each industry. The simulated data is transformed using the square root of the covariance matrix in the same manner as above. Once again, for simplicity, we do not allow second-order cross-correlation.

\subsection{MonteCarlo Simulation of Efficient Frontiers}
For each DGP, we first calculate the true frontier using Equation \ref{eq:EF} which is plotted in black in Figures \ref{fig:Gaussian} and \ref{fig:others}. To demonstrate the rescaling effect of the decision rules and the DGPs, we examine particular portfolios at particular levels of risk-aversion, represented as dots along the frontiers in Figures \ref{fig:Gaussian} and \ref{fig:others}. These levels are selected such that the allocation to the GMV portfolio varies over $[-1, 1]$ in increments of $0.2$ \footnote{Equivalently, the allocation to the MRAR varies over $[0, 2]$. Note that as the allocation the MRAR portfolio exceeds 1, the GMV is short-sold.}.  Additionally, we compare the decision rule against the ``plug-in" rule which uses sample estimators, and a decision rule that uses a two-factor statistical factor model to construct estimates. 

For each DGP, we simulate 10,000 samples of $T=36$ months, representing a time period where the distribution of returns may stay fairly stationary. In each sample, we estimate the mean and covariance parameters using the estimators of the decision rules. Plugging these estimates and the specified risk aversion coefficients into equation \ref{eq:EF} generates the estimated efficient frontiers. We calculate the average frontier over all samples and this is plotted in Figure \ref{fig:Gaussian} and \ref{fig:others}, where each dot corresponds to a portfolio with the specified risk aversion constant. 

Most of the shrinkage estimators described in section \ref{subsec:pdr} require the specification of prior information. For the linear shrinkage method, we use the constant correlation model (CCM) as the shrinkage target. This covariance matrix is constructed from assuming that the correlations between all assets are the same, while individual variances are constructed using sample estimators. The CCM encodes the prior information encompassing the empirical fact that assets are mostly positively correlated. For the Data-and-Model decision rule, we use a two-factor statistical factor model as the investor's prior pricing model. Additionally, the covariance shrinkage methods requires the specification of a mean estimator to construct a portfolio. For this purpose, we use the sample mean estimator. 

% The prior for the nonlinear shrinkage method is given by:

\section{Results}\label{sec:results}
% This approach to portfolio construction is known as the ``plug-in`` approach. 
\begin{figure}[t!]
\centering
\resizebox{\columnwidth}{!}{
\input{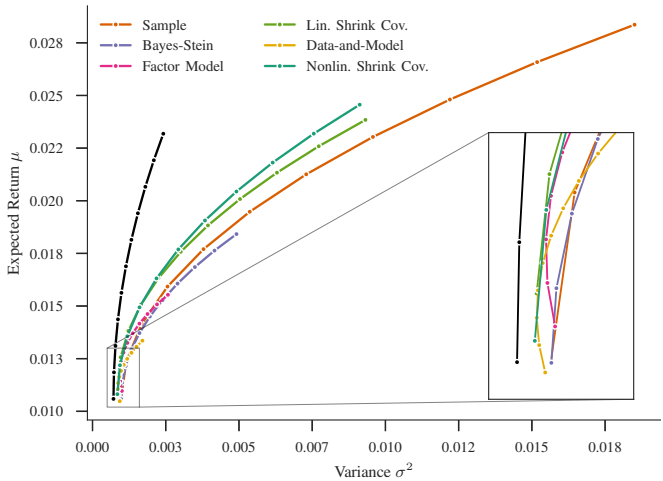}
}
\caption{Efficient Frontiers Under a Gaussian DGP}
\label{fig:Gaussian}
\end{figure}

\begin{figure*}[h!]
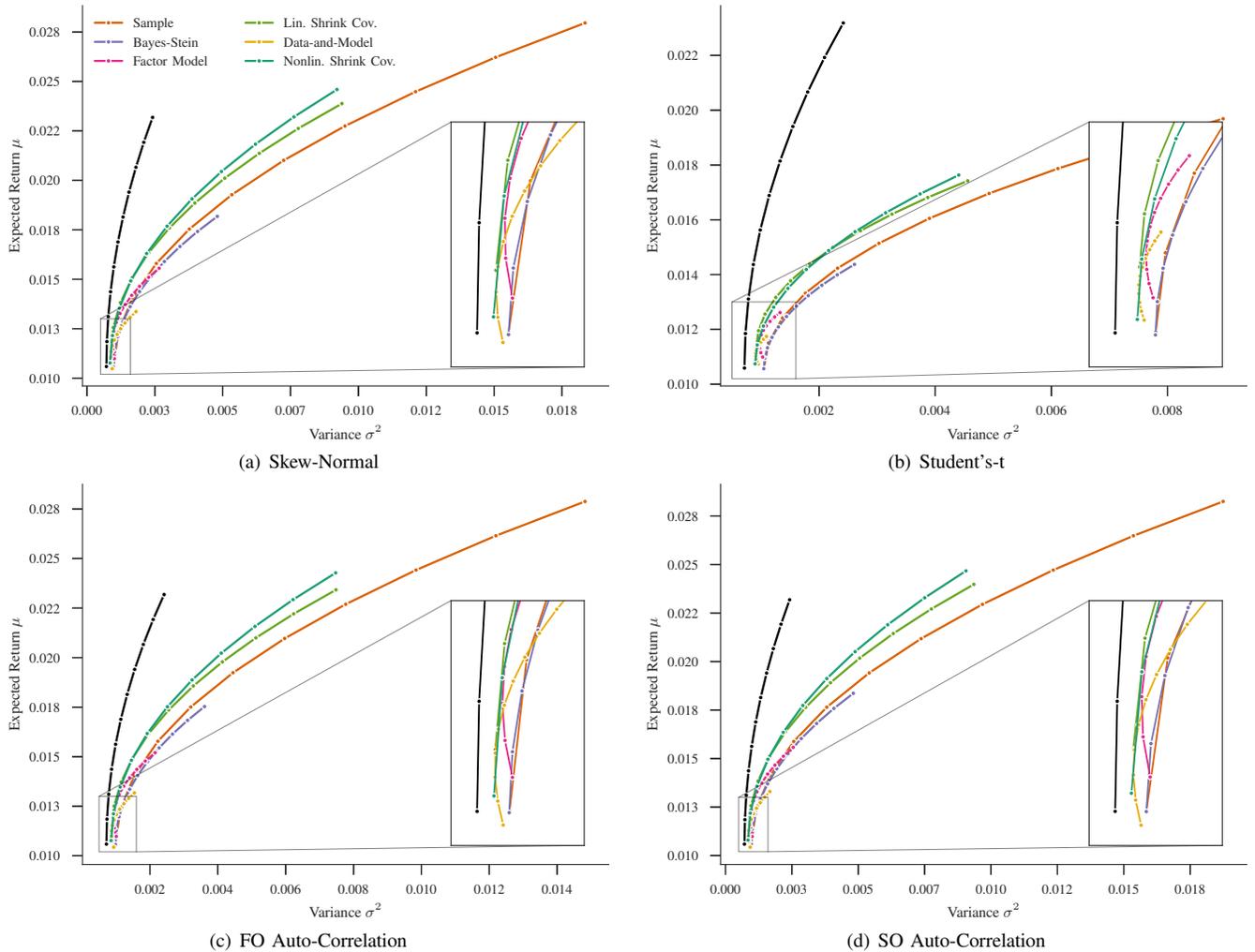

\centering
\subfigure[Skew-Normal]{
\resizebox{0.98\columnwidth}{!}{
\input{pgf/SkewNorm.pgf}
}
}
% \hfil
\subfigure[Student's-t]{
\resizebox{0.98\columnwidth}{!}{
\input{pgf/students-t.pgf}
}
}
% \hfil
\subfigure[FO Auto-Correlation]{
\resizebox{0.98\columnwidth}{!}{
\input{pgf/AR1.pgf}
}
}
% \hfil
\subfigure[SO Auto-Correlation]{
\resizebox{0.98\columnwidth}{!}{
\input{pgf/GARCH.pgf}
}
}
\caption{Efficient frontiers of optimal diversification rules under different distributions.}
\label{fig:others}
\end{figure*}

In the standard decision theoretic framework, a method is evaluated against competitors by their expected mean-variance utility for a fixed level of risk aversion. Considered visually, this means that the \textit{n}\textsuperscript{th} dot on one frontier is compared with the \textit{n}\textsuperscript{th} dot on another frontier in Figure \ref{fig:Gaussian}.  If the method results in a superior expected utility at each level of risk aversion, it is considered a superior method. We call this risk-aversion-specific (RAS) dominance. For example, in Table \ref{tab:expU}, the Bayes-Stein rule has RAS dominance over the plug-in as was reported in the original study ~\cite{jorion1986bayes}.

\begin{table}[t!]
    \centering
    \caption{Loss in expected utility expressed as a percentage of the true optimal utility of each decision rule compared for specified levels of risk aversion $\gamma$. }
    \label{tab:expU}
    \resizebox{\columnwidth}{!}{
    \begin{tabular}{l|rrrrrr}
    \toprule
    $\gamma$ &  Sample &  \makecell[r]{Bayes-\\Stein} &  Factor &  \makecell[r]{Linear\\Shrink\\Cov.} &  \makecell[r]{Data-\\and-\\Model} &  \makecell[r]{Nonlinear\\Shrink\\Cov.} \\
    \bottomrule
    \toprule
    $1.5e8$ &    235.9 &        236.4 &   240.3 &             220.7 &           229.9 &                218.9 \\
    74.0        &    177.7 &        167.2 &   160.5 &             159.1 &           156.3 &                157.8 \\
    37.0        &    161.5 &        131.8 &   119.7 &             132.1 &           117.9 &                130.6 \\
    25.0        &    162.5 &        113.7 &    98.7 &             121.3 &            97.8 &                119.4 \\
    19.0        &    170.1 &        104.5 &    87.9 &             118.2 &            87.0 &                115.8 \\
    15.0        &    180.0 &        100.0 &    82.5 &             118.7 &            81.4 &                115.8 \\
    12.0        &    190.2 &         98.0 &    80.1 &             120.8 &            78.6 &                117.5 \\
    11.0        &    200.0 &         97.4 &    79.3 &             123.7 &            77.4 &                120.1 \\
    9.0         &    209.1 &         97.7 &    79.5 &             126.9 &            77.2 &                122.9 \\
    8.0         &    217.4 &         98.4 &    80.3 &             130.1 &            77.4 &                125.8 \\
    7.0         &    224.9 &         99.3 &    81.3 &             133.1 &            78.1 &                128.6 \\
    \bottomrule
    \end{tabular}
    }
    
\end{table}

In the standard decision theoretic framework, it is possible that a method has RAS-dominance over another, yet has the unappealing and counter-intuitive property that it has worse expected returns for each level of variance. Figure \ref{fig:Gaussian} reveals that decision rules achieve RAS-dominance over the sample frontier largely by ``rescaling" the frontier, to improve the risk-return trade-off for the given risk-aversion. However, if we compare the expected return and risk instead, ie. performance in terms of the entire frontier, we see that the performance improvement is much less, and possibly worse. 

In our analysis, we compare methods by considering the efficient frontier as a whole. Visually, this means that we consider methods with frontiers that lie closer to the true frontier (black) as superior. If one method frontier lies closer to the true frontier than another at all levels of variance, we say that this method has \textit{frontier dominance} over the others. Equivalently, \textit{frontier dominance} occurs if one method has a greater mean return for all levels of variance. For example, in Figure~\ref{fig:Gaussian}, the nonlinear-shrinkage for covariance method has frontier dominance over the other methods, whilst the sample estimators has frontier dominance over the Bayes-Stein method. 

By considering performance in terms of the entire frontier, we see some phenomena in Figure \ref{fig:Gaussian} that are immediately apparent and startling, running counter to some of the prevalent ideas in the portfolio selection literature. In Figure \ref{fig:others}, we see that these findings are robust to deviations from Normality. 

Firstly, considered from the frontier perspective, we see that the plug-in rule with sample estimators has frontier dominance over the Bayes-Stein rule. However, the root mean-square error of the Bayes-Stein estimator  is approximately $64\%$ of that for the sample mean estimator. It is surprising that frontier dominance occurs in spite of the greater predictive performance of the Bayes-stein mean estimator over the sample estimator. 

This result implies that superior out-of-sample portfolio performance can be attained by the plug-in rule by first constructing the efficient frontier, and then treating the risk-aversion parameter as a hyper-parameter to be optimised. This, in fact, is the manner in which mean-variance portfolio selection is commonly carried out in practice \cite{boyd2017multi}. 

Secondly, we see that the nonlinear shrinkage covariance method dominates the remaining frontiers. Here, we see that improving the prediction of the covariance matrix has a greater impact on frontier performance than improving prediction of the mean, even in a low-dimensional setting. This is in contrast to the results of \citet{chopraziemba1993}, who analyse portfolio performance in the standard decision theoretic framework, and conclude that errors in the means are more impactful for portfolio performance than errors in the covariance matrix. 

Lastly, it is worth pointing out that the relative performance of the different rules depends on the level of out-of-sample variance under consideration. Thus, the common practice of selecting a portfolio decision rule and then optimising hyper-parameters afterwards should be extended to consider multiple decision rules simultaneously.  

In Figure \ref{fig:others}, we see that the ranking of the performance between the different decision rules largely remains the same as we depart from the IID Gaussian setting. It appears that changing the DGP from the MVG to a MVT distribution had the largest impact on the shape of the frontier and the relative performance of the decision rules. Note, in particular, that under the MVT the frontier of the nonlinear shrinkage estimator does not dominate the frontier of the linear shrinkage estimator. 

Under the DGP with AR(1) returns, we see that the variance of each portfolio along the frontiers of all the decision rules has decreased. This agrees with the intuition that negative autocorrelation should reduce the variance of the various estimators, and in turn, the variance of portfolios. 

In comparison to the MVG DGP, the efficient frontiers of the decision rules under the GARCH and MVSN DGPs appear to remain almost exactly the same. This is surprising as it is expected that these deviations would affect the predictive accuracy of the mean and covariance estimators, resulting in worse performance. Since the plotted frontiers are the average of a distribution of frontiers, it is not evident if other moments of the frontier distributions have been affected under the GARCH and MVSN DGPs. Inspection of these variances (not shown) appears to show some differences to the MVG GDP, but further analysis is required to determine if these differences are statistically significant.

\section{Conclusion}\label{sec:conclusion}
The performance of seminal portfolio decision rules is not well understood and the value of mean-variance optimisation is currently debated. In this article, we study the efficient frontiers spanned by different decision rules and examine how they are affected by departures to the IID Gaussian setting, which is assumed in their design. Our experimental results suggest that excess kurtosis and autocorrelation can substantially impact performance whilst performance differences resulting from skewness and garch effects appear negligible. 

However, it is important to note a limitation of our study is that we consider only a single set of particular deviations from Normality. Thus, the relative importance of these deviations can not be solely inferred from our results alone. Further analysis is required to determine how performance is affected as the simulation parameters are varied. 

By considering performance in terms of the efficient frontiers, rather than comparing portfolios for fixed levels of risk aversion, we present findings which appear to oppose the prevailing ideas in the literature. Firstly, sample estimator has an efficient frontier that dominates the Bayes-Stein estimator for means. Secondly, improving the covariance matrix seems to have a greater effect on performance than the mean estimator. 

Our results indicate some immediate directions of further research: firstly the performance improvement through hyper-parameter optimisation of decision rules. Secondly, the formalisation of metrics to compare efficient frontiers to determine which components of the decision rules are most impactful in improving performance across the frontier. 

% \nocite*
\bibliographystyle{abbrvnat}
\bibliography{MVdist}

\end{document}